\documentclass[12pt, a4paper]{extarticle}
\usepackage[top=1cm,width=17cm,height=27cm]{geometry}
\usepackage[utf8]{inputenc}
\usepackage[T1]{fontenc}
\usepackage{lmodern}
\usepackage{tabularx}
\usepackage{mathtools}
\usepackage{etoolbox}
\usepackage[mathscr]{eucal}
\let\savemathscr=\mathscr
\usepackage{mathrsfs}

\let\mathscr=\savemathscr
\let\savemathscr=\relax
\let\savermdefault=\rmdefault
\let\savefamilydefault=\familydefault
\let\savebfdefault=\bfdefault
\usepackage{tgchorus}
\renewcommand{\rmdefault}{\savermdefault}
\fontfamily{\savefamilydefault}
\renewcommand\bfdefault{\savebfdefault}
\DeclareMathAlphabet{\gcal}{T1}{qzc}{m}{n}
\usepackage{MnSymbol}
\usepackage[svgnames]{xcolor}

\usepackage{tikz}
\usetikzlibrary{trees, automata, positioning, arrows}
\tikzset{
  color = Navy!100, 
  node distance = 5em,
  every state/.style = {color = Navy!100,
    minimum size = 1.5em,
    inner sep = 1pt},
  initial text={},
  initial distance={1em},
  accepting/.style = {accepting by arrow},
  accepting distance={1em},
  every edge/.style={
    draw,
    color = Navy!100, 
    ->,>=stealth,
    auto},
  every pin edge/.style={
    draw,
    color = Navy!100, 
    ->,>=stealth,
    auto}}
\usepackage{authblk}
\usepackage{enumitem}
\setlist[itemize]{noitemsep, topsep=1pt}
\let\setminus=\smallsetminus
\newcommand{\bbbn}{{\mathrm{I\!N}}}
\newcommand{\bbbz}{{\mathchoice {\hbox{$\mathsf\textstyle Z\kern-0.4em Z$}}%
                                {\hbox{$\mathsf\textstyle Z\kern-0.4em Z$}}%
                                {\hbox{$\mathsf\scriptstyle Z\kern-0.3em Z$}}%
                                {\hbox{$\mathsf\scriptscriptstyle%
                                Z\kern-0.2em Z$}}}}

\newcommand{\bbbr}{{\mathrm{I\!R}}}

\newlength{\wpdpth}
\newcommand{\pset}{\raisebox{\wpdpth}{\large$\wp$}}

\newcommand{\dist}{\mathsf{Dist}}
\newcommand{\bfr}{\mathbf{r}}

\newcommand{\bfS}{\mathbf{S}}
\newcommand{\bfone}{\mathbf{1}}
\newcommand{\calT}{\mathcal{T}}
\newcommand{\calE}{\mathcal{E}}
\newcommand{\quot}{\raisebox{-.2ex}{$/$}\raisebox{-.4ex}{$\equiv$}}
\numberwithin{equation}{section}
\title{Signal automata and hidden Markov models}
\author{Teodor Knapik}
\affil{ISEA,  Universit\'e de la Nouvelle Cal\'edonie}
\begin{document}
\everymath{\color{Navy}}
\everydisplay{\color{Navy}}
\maketitle
\begin{abstract}
  A generic method for inferring a dynamical hidden Markov model from
  a time series is proposed. Under reasonable hypothesis, the model is
  updated in constant time whenever a new measurement arrives.
\end{abstract}
\section{Introduction}
\label{sec:intro}
\nocite{higuera:gram,verwer:compet}
Many natural dynamical systems are poorly understood due to their
overwhelming complexity. Their scale
is often large and leaves almost no area for experimental study.
Predicting how the current trajectory will evolve in the future is
a challenging task, especially when no other trajectory is known.
Indeed, as such systems cannot be reset to an initial state and restarted,
at most one trajectory can be observed. 
Any ecosystem, the climate and, many social and geological
phenomena evolving along their unique trajectory are typical examples
where predictive modelling is highly nontrivial. Among many approaches
for addressing this challenge, hidden Markov models (HMM) gained
popularity in related research areas (see e.g.\@
\cite{mcclintock:ecoHMM,holsclaw:rainHMM,mathew:mobilityHMM,
  nawaz:seismicHMM}).

Introduced in \cite{baum-petrie:inference,baum-eagon:estim}, hidden
Markov models formalise, among others, the idea that a dynamical
system evolves through its state space which is unknown to an observer
or too difficult to characterise - only a few attributes are known and
can be observed among a huge number of mostly unknown (hidden) attributes
governing the dynamics of the system. A hidden Markov model (HMM)
possesses a set of states and its evolution may be seen as sequence
categorical distributions over that set while satisfying the Markov
property, exactly like a Markov chain does.  In addition, each state
has its associated distribution over the set of possible
observations. Thus, an HMM can be seen as a device which, evolving
from its initial distribution over states, produces a sequence of
distributions over observations. Estimating the accuracy of an HMM
with respect to the dynamical system, the HMM is supposed to model, is
easy: one uses the HMM to compute the probability of the observed
sequence of measurements of the trajectory of the system. Producing an
accurate HMM given a time series of observations of the system is much
more challenging. The first, and until today, the most popular
algorithm inferring an HMM from a time series is the Baum-Welch
algorithm \cite{baum:baum-welch,welch:baum-welch}. Since the seminal
tutorial of Rabiner \cite{rabiner:HMM}, there has been a significant
number of variations on the theme of the Baum-Welch algorithm. Among
them, several successful proposals of its on-line extensions have been
published: \cite{baldi-chauvin:on-line}, \cite{mongillo:on-line} or
\cite{chris-harrisson:on-line}, to mention at least a few. The present
note breaks with that tradition by bringing a theoretical computer
scientist's yet another point of view.

The main motivation of this work is the predictive modelling of
dynamical systems where the laws governing their dynamics are
difficult to put in a general form and where only one trajectory can
be observed. The observations come form measurements made at regular
time intervals, thus producing a time series, also called a
signal. The model is built from data collected so far and used for
forecasting the values of future measurements. When the present
becomes past and a new measurement becomes available, the existing
model is updated. The algorithmic complexity of updating the model
is expected to be in $o(f(n))$ (ideally in $O(1)$) when $f(n)$ is the
complexity of its building from scratch.

Signal automata introduced here are generic ``syntactic''
devices. Their ``semantics'' is given in term of HMMs. To exist, a
signal automaton needs a state-producing plugin which may be
application-specific. One can imagine the plugin as a procedure which
extracts meaningful patterns from the signal, makes a lossy
compression of the signal, filters it or classifies it according to
some criteria acquired throughout unsupervised or supervised
learning. The definition of signal automaton does not rely on specific
plugin but rather provides a generic wrapper.  The set of known states
of a signal automaton and transitions between those states evolve in
time. Whenever a new measurement is made, the plugin, applied to the
whole signal available so far, returns a state of the automaton which
becomes its current state. It may be one of previously visited states
or a new one. In both cases, the new measurement affects transitions
of the automaton, either by adding a new transition or by altering an
existing one. Two other plugins are needed to build the HMM
corresponding to the current automaton. The natures of the latter
plugins are very different from the former. All plugins are supposed
to have parameters. Searching within the parameter space may be used
for fitting the model.  The resulting HMM is then used for
forecasting in the usual way. A bit less standard lookahead
forecasting is also discussed here.

\section{Signal automata}
\label{sec:saut}
A prefix $s_0s_1\ldots s_n$
of a sequence $(s_i)_{i\in\bbbn}$ is written $s_{\le n}$ and a portion
$s_ms_{m+1}\ldots s_n$ of it, for $m\le n$, is written
$s_{m\,:\,n}$.

A sequence of observations $\bfr:=(r_i)_{i\in\bbbn}$ of a dynamical
system is assumed to range over some set $R$. Observations arrive at
regular time intervals numbered $0, 1, 2,\ldots$.  In the sequel,
cases where $R$ is a normed vector space or a metric space are
considered.  At time $n$, only prefix $r_{\le n}$ of $\bfr$ is known
and is called a \emph{signal at $n$}.  A map
$\kappa^*_\tau\colon R^*\to C$, called a \emph{classifier}, from a set
$R^*$ of finite sequences over $R$ into a set $C$ is the main plugin
of a signal automaton defined in the sequel. Which kind of set $C$
should be considered is not discussed here. It cloud be e.g. a set of
meaningful patterns appearing in $\bfr$, a set of highly compressed
signals, a set of clusters of signals, or a set of some averages of
the signal.  The classifier depends on a parameter tuple $\tau$
varying within some finite-dimensional space of parameters $T$. The
\emph{compression factor of $\bfr$ by $\kappa^*_\tau$} is a function
$\xi\colon\bbbn\to\bbbr_+$ such that the map
$n\mapsto|\kappa^*_\tau(\{r_{\le i}\mid i\le n\})|$ is in
$\theta(\xi(n))$ (in the sense of asymptotic behaviour).  A classifier with
the compression factor in $o(n)$ is preferred.  A \emph{precursor} of
a classifier $\kappa^*_\tau$ is a map
$\kappa_\tau\colon C\times R\to C$ such that, for every signal
$r_{\le k}$, one has
$\kappa^*_\tau(r_{\le k}) = \kappa_\tau(\kappa^*_\tau(r_{\le k-1}),
r_k)$. If $\kappa^*_\tau$ has a precursor, the time complexity of
computing $\kappa^*_\tau(r_{\le k})$, after previously computing
$c_{k-1}=\kappa^*_\tau(r_{\le k-1})$, does not depend on $k$ as it is
reduced into computing $\kappa_\tau(c_{k-1}, r_k)$. For instance, a
family of classifiers having precursors computing in constant time can
be defined using exponential averages, assuming that basic arithmetic
operations are computed in constant time on fixed size floating point
data type.

A \emph{statistical function}
$\sigma_\tau\colon R^*\times\pset_{\text{fin}}(\bbbn)\to\bbbr_+$ computes
some relevant statistics. It depends on the same parameter tuple
$\tau$ as does a classifier. Statistical value
$\sigma^*_\tau(r_{\le i},\{i_1,\ldots, i_k\})$ is computed from instants
$\{i_1,\ldots, i_k\}\subseteq\{0,1,\ldots, i\}$ and observations
$r_{i_1},\ldots,r_{i_k}$ although the latter may in some applications
remain unused.
For instance, $\sigma^*_\tau$ may yield a mean value of $r_{i_1},\ldots,r_{i_k}$,
count how many observations among
$r_{i_1},\ldots,r_{i_k}$ belong to some specific region of $R$ or
what is the latest occurrence $i_j$, for $1\le j\le k$, of an observation
in that region. Another example is a discounted counting with respect
to a current instant, say $n$:
$\sigma^*_\tau(\_,\{i_1,\ldots, i_k\})=k-\sum_{i\in\{i_1,\ldots, i_k\}}\delta^{n-i}$
where $0\le \delta<1$ is one of parameters of $\tau$. The idea is that
older events count less than newer ones.
Like for a classifier, a \emph{precursor} of a statistical function
$\sigma^*_\tau$ is is a map
$\sigma_\tau\colon\bbbr_+\times R\times\pset_{\text{fin}}(\bbbn)\to\bbbr_+$
such that, for every $\{i_1\ldots i_k\}\subseteq\{0,1,\ldots, i\}$,
one has
\par\noindent\hfil
$\sigma^*_\tau(r_{\le k},\{i_1\ldots i_k\}) =
\sigma_\tau(\sigma^*_\tau(r_{\le k-1}, \{i_1,\ldots,
i_{k-1}\}),r_k,i_k)\enspace.$
\smallskip

Consider a classifier $\kappa^*_\tau$. A \emph{signal automaton} $\bfS$
defined by $(\bfr,\kappa^*_\tau)$ 
is a sequence $\bfS:=(S_i)_{i\in\bbbn}$. A term $S_i$ of that sequence
is called an   \emph{instantaneous signal automaton at $i$} (\emph{ISA} for short). It is a tuple $S_i:=(Q_i,\iota_i,\Theta_i)$ where
\begin{itemize}
\item $\iota_i := \kappa^*_\tau(r_{\le i})$ is its \emph{current state},
  in particular $\iota_0=\kappa^*_\tau(r_0)$,
\item $Q_i$  is the set of \emph{states} of $S_i$ defined inductively by
  \begin{itemize}
  \item $Q_0:=\{\iota_\bot,\iota_0\}$, where
    $\iota_\bot\notin\kappa^*_\tau(\{r_{\le n}\mid n\in\bbbn\})$,
  \item $Q_i:=Q_{i-1}\cup\{\iota_i\}$, for $i>0$,
  \end{itemize}
\item $\Theta_i\colon Q_i\times Q_i \to \pset\{0,1,\ldots, i\}$ is its
  \emph{instants matrix} defined inductively by
  \begin{itemize}
  \item $\Theta_0(\iota_\bot,\iota_\bot) := \emptyset$,
    $\Theta_0(\iota_\bot,\iota_0) := \{0\}$,
    $\Theta_0(\iota_0,\iota_\bot) := \emptyset$,
    $\Theta_0(\iota_0,\iota_0) := \emptyset$,
  \item case $\iota_i\in Q_{i-1}$ (viz., $Q_i = Q_{i-1}$) with $i>0$
    \par
    $\begin{array}{@{}lcll@{}}
      \Theta_i(p,q) &:=& \Theta_{i-1}(p,q),
      & \text{for }(p,q)\in Q_i\times Q_i \setminus{(\iota_{i-1},\iota_i)},\\
       \Theta_i(\iota_{i-1},\iota_i)
                    &:=&\Theta_{i-1}(\iota_{i-1},\iota_i)\cup\{i\},
    \end{array}$
  \item case $\iota_i\notin Q_{i-1}$ (viz., $\iota_i$ is a \emph{new state})
    with $i>0$
    \par
    $\begin{array}{@{}lcll@{}}
      \Theta_i(p,q) &:=& \Theta_{i-1}(p,q),
      & \text{for }(p,q)\in Q_{i-1}\times Q_{i-1}\\
      \Theta_i(\iota_i,q) &:=& \emptyset,
      & \text{for }q\in Q_i,\\
      \Theta_i(p,\iota_i) &:=& \emptyset,
      & \text{for }p\in Q_{i-1}\setminus\{\iota_{i-1}\},\\
      \Theta_i(\iota_{i-1},\iota_i) &:=&\{i\}.
    \end{array}$
  \end{itemize}
\end{itemize}
The inductive step of this definition may be seen as computing
$S_i\gets\mathsf{nextISA}(\kappa^*_\tau,r_{\le i},S_{i-1})$
by means of some implemented function $\mathsf{nextISA}$.
If $\kappa^*_\tau$ has a precursor, a variant
of $\mathsf{nextISA}$ is used:
$S_i\gets\mathsf{nextISA}(\kappa_\tau,\iota_{i-1},r_i,S_{i-1})$.
The complexity of $\mathsf{nextISA}$ is determined
by the complexity of $\kappa_\tau^*$ or $\kappa_\tau$. The
latter complexity is denoted by $\eta(n)$.
If $\kappa_\tau(\iota_{n-1},r_n)$ is computed in constant time
then $\mathsf{nextISA}(\kappa_\tau,\iota_{n-1},r_n,S_{n-1})$ can be computed in
almost constant time provided appropriate implementations
of sets $Q_n$ and $\Theta_n(p,q)$, and of the sparse matrix
$\Theta_n$.

In the sequel, the complexity is discussed under the assumption that a
precursor of a classifier is used. It is also relevant to distinguish
between two situations.  At present time, say $n$, one may need to
compute $S_n$ from scratch using available observations. One speaks
then of \emph{computing ISA at $n$} and it is done in time
$O(n\eta(n))$. If however $S_{n-1}$ has been already computed, one
speaks of \emph{updating ISA at $n$}, which is done in time
$\theta(\eta(n))$.

A signal automaton may be seen as an automaton with evolving state
space, changing its current state according to the arriving new
observation. Each transition $\Theta_i(p,q)$ records time points at
which a move from $p$ to $q$ occurred. There are two possible
situations for a current state.  Either the state has never been
encountered and then it has no outgoing transition, viz.,
$\cup_{q\in Q_i}\Theta_i(\iota_i,q)=\emptyset$, or it has been
previously encountered and it has at least one outgoing transition. In
the former situation one speaks of a \emph{new state}. Note that if
$S_i$ has a state with no outgoing transition, it is necessarily its
current state $\iota_i$.

\section{From signal automata to hidden Markov models}
\label{sec:hmm}
A hidden Markov model (HMM for short) $M=(E, Q, \alpha, \calT, \calE)$
consists of
\begin{itemize}
\item a set $E$ of (observable) events,
\item a set $Q$ of states,
\item an initial distribution $\alpha\in\dist(Q)$ of states,
  where $\dist(Q):=\{\alpha\in [0,1]^Q\mid \sum_{q\in Q}\alpha(q)=1\}$ stands
  for the set of (categorical) distributions over $Q$, 
\item a transition matrix $\calT\colon Q\times Q\to [0,1]$ which is
  a (right) stochastic matrix (its every row is in $\dist(Q)$),
\item an emission matrix $\calE\colon Q\times E\to [0,1]$, each row
  of which is in $\dist(E)$.
\end{itemize}
The use of an emission matrix is appropriate only when $E$ is a
discrete space. In the continuous case, $\calE$ is a map
$E\colon Q\to E\to [0,1]$ such that $E(q)\colon E\to[0,1]$ is
a probability density function for each state $q\in Q$. In other
words, $\calE$ is a vector over $Q$ of probability density functions
over $E$.

To turn an ISA $S_i=(Q_i,\iota_i,\Theta_i)$
defined by $(\bfr,\kappa^*_\tau)$ into an HMM with a discrete space 
of events one needs to cluster $R$ by means of an equivalence relation
``$\equiv$'' such that  the quotient space $R\quot$ is discrete.
Let $C_i=\{c\in R\quot\mid \{r_i\mid i\le n\}\cap c\neq\emptyset\}$.
One needs also a pair statistical functions $(\sigma^*_\tau,\rho^*_\tau)$.
Then HMM $M_i=(E_i, Q_i', \alpha_i, \calT_i, \calE_i)$ is obtained from
$S_i$ by taking
\begin{itemize}
\item $E_i:=C_i\cup\{r_\emptyset\}$, where $r_\emptyset\notin R\quot$,
\item $Q_i':=Q_i\setminus\{\iota_\bot\}\cup\{q_\emptyset\}$ with an
  additional state $q_\emptyset\notin\kappa^*_\tau(R^*)\cup\{\iota_\bot\}$,
\item $\alpha_i:=\bfone_{\iota_i}$ is the indicator function of $\iota_i$
  over $Q_i'$,
\item 
  $\begin{array}[t]{@{}rcl@{}}
     \calT_i(q_\emptyset,q_\emptyset)&:=&1,\\
     \calT_i(q_\emptyset,q)&:=&0,\quad\text{for }q\in Q_i\setminus\{\iota_\bot\},\\
     \calT_i(p,q_\emptyset)&:=&1\quad\text{if }
                              \bigcup_{q\in Q_i}\Theta_i(p,q)=\emptyset,
                              \quad\text{for }p\in Q_i\setminus\{\iota_\bot\},\\
     \calT_i(p,q_\emptyset)&:=&0\quad\text{if }
                           \bigcup_{q\in Q_i}\Theta_i(p,q)\neq\emptyset,
                                \quad\text{for }
                                p\in Q_i\setminus\{\iota_\bot\},\\[1ex]
     \calT_i(p,q)&:=&\displaystyle\frac{\sigma^*_\tau(r_{\le i},\Theta_i(p,q))}
                      {\sum_{s\in Q_i\setminus\{\iota_\bot\}}\sigma^*_\tau(r_{\le i},
                      \Theta_i(p,s))},
                      \quad\text{for }p\in Q_i,  q\in Q_i\setminus\{\iota_\bot\},
  \end{array}$
\item 
  $\begin{array}[t]{@{}rcl@{}}
     \calE_i(q_\emptyset,r_\emptyset)&:=&1,\\
     \calE_i(q_\emptyset,c)&:=&0,\quad\text{for }c\in C_i,\\[1ex]
     \calE_i(q,c)&:=&\displaystyle
     \frac{\rho^*_\tau(r_{\le i},\{j\in \Theta_i(p,q)\mid
                      r_j\in c,p\in Q_i\})}
                      {\rho^*_\tau(r_{\le i},\bigcup_{p\in Q_i}
                      \Theta_i(p,q))},\quad
                      \text{for }q\in Q_i\setminus\{\iota_\bot\}
                      \text{ and }c\in C_i.
  \end{array}$
\end{itemize}
Clustering $R$ may be computationally expensive but once done
properly, finding the cluster of a given observation is done in constant
time. A reasonable assumption about complexities of
$\sigma^*_\tau(r_{\le i},\{i_1,\ldots, i_k\})$ and
$\rho^*_\tau(r_{\le i},\{i_1,\ldots, i_k\})$ is that, provided
an appropriate implementation of $\bfr$, both depend on only $k$,
and depend on it linearly. Under this assumption, a direct computation
$M_n\gets\mathsf{ISAtoHMM}(\sigma^*_\tau,\rho^*_\tau,r_{\le n},S_n)$ from $S_n$,
by means of some implementation $\mathsf{ISAtoHMM}$
can be done in time $\theta(n)$ when $\calT_n$ and $\calE_n$ have
an appropriate implementation as sparse matrices. Indeed, the family
$\{\Theta_n(p,q)\mid p,q\in Q_n\}$ consists of pairwise disjoint sets.
Thus, the sum of computation times of all cells of $\calT_n$ and
$\calE_n$ is in $\theta(n)$. Moreover, $n\mapsto |E_n|\in O(n)$.

One can also compute
$M_n\gets\mathsf{nextHMM}(\sigma^*_\tau,\rho^*_\tau,r_{\le n},S_n,M_{n-1})$
$M_n$ by updating $M_{n-1}$, if available,
by means of some implementation $\mathsf{nextHMM}$. This can
be done in constant time under the same assumptions as for
$\mathsf{ISAtoHMM}$.

Providing a pair of $(\sigma^*_\tau,\rho^*_\tau)$ for $(S_i)_{i\in\bbbn}$
may be understood, similarly to an interpretation in mathematical
logic, as assigning a meaning to the signal automaton
defined by $(\bfr,\kappa^*_\tau)$. The associated meaning is the sequence
$(M_i)_{i\in\bbbn}$ of HMMs, each HMM $M_i$ obtained from ISA $S_i$ as
described above. As forecasting is the main motivation of introducing
signal automata,
special care is needed in handling new states. Recall that a state of
$S_i$ is new if it is not a state of $S_{i-1}$ and that it is
necessarily the current state of $S_i$, namely $\iota_i$. In such a
situation, $M_i$ can evolve from its initial distribution $\alpha_i$ only
into its absorbing dummy state $q_\emptyset$ where the only possible
event is dummy event $r_\emptyset$. The frequency of such situations is
given by the ratio $\xi(n)/n$.

\subsection*{Case of continuous event space}
If $R$ is a continuous vector space, one may wish to turn ISA
$S=(Q,\iota,\Theta)$ into an HMM acting directly over $R$ instead of
clustering $R$ into a discrete space. Like in the latter discrete case,
one needs a pair $(\sigma^*_\tau,\rho^*_\tau)$ where $\sigma^*_\tau$ is a
statistical function. However, instead of being a statistical function,
$\rho^*_\tau$ should be a (multivariate) kernel function, like
e.g.\@ multivariate normal kernel:
\par\noindent\hfil
$\phi_H(x)=(2\pi)^{-d/2}|H|^{-1/2}\exp(-\frac{1}{2}x^TH^{-1/2}x)$
\par\noindent
where $d=\mathrm{dim}(R)$ and $H$ is a $d\times d$ matrix, called
a bandwidth matrix and playing a role similar to a covariance
matrix. For any kernel function $\rho^*_\tau$ used here, a bandwidth
matrix is a component of parameters tuple $\tau$.

One defines the associated HMM $M=(E, Q', \alpha, \calT, \calE)$ by
taking $Q'$, $\alpha$ and $\calT$ like in discrete case, and
\begin{itemize}
\item $E:=R\cup\{r_\emptyset\}$, where $r_\emptyset\notin R$,
\item 
  $\begin{array}[t]{@{}rcl@{}}
     \calE(q_\emptyset,x)&:=&\delta_{r_\emptyset}(x)
     \quad(\text{the Dirac delta centred at }r_\emptyset),\\
     \calE(q,r_\emptyset)&:=&0\quad\text{for }
                              q\in Q\setminus\{\iota_\bot\},\\[1ex]
     \calE(q,x)&:=&\displaystyle\frac{1}{|\bigcup_{p\in Q}\Theta(p,q)|}
                    \sum_{p\in Q}\sum_{\makebox[3em][r]{$\scriptstyle j\in\Theta(p,q)$}}
                    \rho^*_\tau(x-r_j)\quad
                    \text{for }q\in Q\setminus\{\iota_\bot\}\text{ and }x\in R,
  \end{array}$
\end{itemize}
The complexities of continuous analogues of $\mathsf{ISAtoHMM}$
and $\mathsf{nextHMM}$, say $\mathsf{ISAtoHMMc}$
and $\mathsf{nextHMMc}$ are like in the discrete case.

The expression of $\calE(q,x)$ is a direct adaptation of the usual
density estimate \cite{rosenblatt:density,parzen:density}, or rather
its multivariate extension \cite{simonoff:smoothing}, using
$\rho^*_\tau$ as kernel function which in turn involves bandwidth
matrix $H$. The latter is essential for the accuracy of kernel density
estimation but methods for finding optimal $H$ are often
computationally expensive.  More recently an objective data-driven
approach for probability density estimation has been introduced
\cite{bernac:density}. Based upon the latter work, fast methods for
computing probability densities have been developed in
\cite{obrien:fastKDE}. These provide a better alternative for
implementing $\calE(q,x)$ than its above expression.  Thus, the latter
should understood as an example rather than a definition.  In other
words, one can plug here the best available algorithm for density
estimation.

\section{Forecasting for finite horizons}
\label{sec:fore}
At instant $i$, only ISA $S_{\le i}$ are known.
Let $M_i=(E_i, Q_i', \alpha_i, \calT_i, \calE_i)$ be a HMM associated
with ISA $S_i$. A \emph{finite horizon} is a natural number $h\in\bbbn$.
A \emph{forecast for horizon $h$ at instant} $i\in\bbbn$ is a sequence
$(f_{i,j})_{1\le j\le h}$ of categorical distributions, or, in the continuous
case, of probability density functions, where
$f_{i,j}=\alpha_i\calT_i^j\calE_i$.

It should be noted that, whenever $\iota_i$ is a new state of $S_i$,
the only forecast for horizon $h\in\bbbn\setminus\{0\}$ is the
sequence with terms $r_\emptyset$ solely, represented by $h$ times
repeated $\delta_{r_\emptyset}$ in the continuous case or an analogous
categorical distribution in the discrete case. This kind of dummy
forecast simply means that no forecast is possible. In other words,
the classifier could not classify the current situation as similar to
some past situation. Thus, such new situation has an unpredictable
future.

With an appropriate implementation of sparse matrices $\calT_n$ and
$\calE_n$, forecasting at time $n$ for horizon $h$ can be done in time
$\theta(h\xi(n))$. In the discrete case, finding $f_{n,j}(c)$ of a
given class $c\in C_n$ can be done in constant time. However, in the
continuous case, computing $f_{n,j}(r)$ for a given value $r\in R$
requires time in $O(n\xi(n))$.

\subsection*{Lookahead forecasting}
In many applications, data analysis starts only at time $n$ after
collecting a substantial amount of observations $r_{\le n}$.  As for
every instant $i\le n-h$, the observed future of the dynamical system
is known at least up to horizon $h$, it makes sense with regard to
forecasting at $h$ to compute each state $\iota_i$ of the signal
automaton, for $i\le n-h$, using not only the observed past
$r_{\le i}$ but more importantly the future already known
$r_{i+1\,:\,i+h}$.  A variant of classifier, called a \emph{classifier
  with $h$-lookahead}, $\kappa^*_{\tau,h}\colon R^*\times R^h\to D$,
is introduced for that purpose. It differs from the formerly
considered classifier only by its requiring a second
argument of length $h$. The most straightforward example of a
classifier with $h$-lookahead is a map which given $r_{\le i}$ and
$r_{i+1\,:\,i+h}$ returns $c_{i+1\,:\,i+h}$,  where $c_j=[r_j]_\equiv$,
or more exactly, a symbolic representation of it. In particular,
when $R\quot$ is finite, it may be considered as an alphabet and
$c_{i+1\,:\,i+h}$ as a word over that alphabet.

From now on, let $(S_i)_{i\in\bbbn}$ denote the signal
automaton defined by $(\bfr,\kappa^*_{\tau,h})$ where $\kappa^*_{\tau,h}$
is a classifier with $h$-lookahead and $(M_i)_{i\in\bbbn}$ be the
associated sequence of HMMs obtained from $(S_i)_{i\in\bbbn}$ using
a pair $(\sigma^*_\tau,\rho^*_\tau)$. The lookahead does not
impact procedures $\mathsf{nextISA}$, $\mathsf{ISAtoHMM}$,
$\mathsf{nextHMM}$, $\mathsf{ISAtoHMMc}$ and $\mathsf{nextHMMc}$.
Their use for computing $(S_i)_{i\in\bbbn}$ and $(M_i)_{i\in\bbbn}$ remains
unchanged. However, at present time $n$, only $S_{\le n-h}$
and $M_{\le n-h}$ can be computed in that way. Beyond
instant $n-h$, some estimated observations, say $\hat r_{n+1\,:\,i+h}$,
are required.

The \emph{$i$-th ISA at $n$ with $h$-lookahead}
$\hat S_{n,i}=(\hat Q_{n,i},\hat\iota_{n,i},\hat\Theta_{n,i})$ together
with the associated \emph{$i$-th HMM at $n$ with $h$-lookahead}
$\hat M_{n,i}=(\hat E_{n,i},\hat Q_{n,i}',\hat\alpha_{n,i},\hat\calT_{n,i},
\hat\calE_{n,i})$ is defined inductively as follows.
For $i\le n-h$, define $S_{n,i}:=S_i$ and $\hat M_{n,i}:=M_i$.
For $n-h<i\le n$, let $\hat r_{i+h}$ be obtained by sampling from
$\hat f_{n,i-1,h+1}:=\hat\alpha_{n,i-1}\hat\calT_{n,i-1}^{h+1}\hat\calE_{n,i-1}$
and let
$\hat\iota_{n,i}:=\kappa^*_{\tau,h}(r_{\le i}, r_{i+1\,:\,n}\hat r_{n+1\,:\,i+h})$.
Assuming that $\hat\iota_{n,i}$ is not a new state, $\hat S_{n,i}$ and
$\hat M_{n,i}$ are defined like $S_i$ and $M_i$ but using $\hat\iota_{n,i}$
instead of $\iota_i$ and $r_{\le n}\hat r_{n+1\,:\,i+h}$ instead of
$r_{\le i+h}$.

Procedures $\mathsf{ISAtoHMM}$ and $\mathsf{nextHMM}$
work like without lookahead whereas the adequate variant
of $\mathsf{nextISA}$ requires both genuine observations and estimated
observations to be passed to $\kappa^*_{\tau,h}$:
\par\vspace*{-2ex}\noindent\hfil
$\begin{array}[t]{@{}rcl@{}}
    \hat S_{n,i}&\gets&
    \mathsf{nextISA}(\kappa^*_{\tau,h},r_{\le n}\hat r_{n+1\,:\,i+h},\hat S_{n,i-1}),\\
    \hat M_{n,i}&\gets&
    \mathsf{ISAtoHMM}(\sigma^*_\tau,\rho^*_\tau,r_{\le i},\hat S_{n,i}),\\
    \text{or }\hat M_{n,i}&\gets&\mathsf{nextHMM}(\sigma^*_\tau,\rho^*_\tau,
                                  r_{\le i},\hat S_{n,i},\hat M_{n,i-1}).
 \end{array}$
 \par\noindent
Their time complexities are the same as without lookahead, so are those
of $\mathsf{ISAtoHMMc}$ and $\mathsf{nextHMMc}$, including all constant-time
precursor variants. The time complexity of sampling $\hat r_{i+h}$
from distribution $\hat f_{n,i-1,h+1}$ is in $O(n)$.

Note that if $\hat\iota_{n,i}$ is a new state,
$\hat S_{n,i+1}, \hat S_{n,i+2},\ldots, \hat S_{n,n}$ are undefined and so
are $\hat M_{n,i+1}$, $\hat M_{n,i+2},\ldots, \hat M_{n,n}$. Lookahead
forecasting
is impossible in such situations. The latter occur with frequency
$h\xi(n)/n$.

Lookahead predictive modelling requires also another operation.
At time $n+1$, new observation $r_{n+1}$ is available. Then the
replacement has to be made according to the following scheme,
where ``$\mapsto$'' means ``is replaced by'':
\par\noindent\hfil
$\begin{array}[t]{@{}r!{\mapsto}l!{\qquad}r!{\mapsto}l@{}}
    \hat S_{n,n-h+1}&S_{n-h+1},&\hat M_{n,n-h+1}&M_{n-h+1},\\
    \hat S_{n,n-h+2}&\hat S_{n+1,n-h+2},&\hat M_{n,n-h+2}&\hat M_{n+1,n-h+2},\\
    \multicolumn{1}{@{}r!{\vdots\ }}{}&&\multicolumn{1}{r!{\vdots\ }}{}&\\
    \hat S_{n,n}&\hat S_{n+1,n},&\hat M_{n,n}&\hat M_{n+1,n}.
 \end{array}$
\par\noindent
In the case where estimated value matches the observed value, viz.,
$\hat r_{n+1}=r_{n+1}$, all ``$\mapsto$'' above become ``$=$''.
If not, $S_{n-h+1}, \hat S_{n+1,n-h+2},\ldots, \hat S_{n+1,n}$ and
$M_{n-h+1}, \hat M_{n+1,n-h+2},\ldots, \hat M_{n+1,n}$ have to be computed.
Also, in any case, $\hat S_{n+1,n+1}$ and $\hat M_{n+1,n+1}$ have to be computed.
If $\iota_{n-h+1}=\hat\iota_{n,h-h+1}$, a few of those computations can
be avoided but the time complexity remains in $\theta(h)$ in the best
scenario, viz., using $\mathsf{nextHMM}$ or $\mathsf{nextHMMc}$ and
precursors computing in constant time.

\section{Conclusion}
\label{sec:conc}
The Baum-Welch and similar algorithms build an HMM with a given number $N$
of states. As the states of the modelled dynamical system are unknown,
finding an adequate $N$ for the model is not obvious.  The approach of
this note bypasses the problem of estimating $N$ by introducing the
concept of signal automaton with a dynamical state space. The author
believes that it is relevant to make this method known to the
scientific community for experimental or theoretical assessment of its
accuracy.

The state space of a signal automaton grows by introducing new
states. As a new state has no outgoing transition, no forecast is
possible when the automaton is in such a state. This is clearly a drawback.
However, in many applications, refraining from forecasting in some
situations may be preferred over providing a bad forecast. Indeed,
when the current situation is very different from all past situations,
it may be accepted that there is no sufficient knowledge to make
predictions.  This is the rationale behind the choice of leaving a new
state with no outgoing transition. As a new measurement arrive, latest
new state is eventually connected to some state, so that, for any
instant $i$, the ISA at $i$ has at most one state with no outgoing
transition.

The fitting of the model can be performed by varying parameters of $\tau$
which intervene in $\kappa^*_\tau$, $\sigma^*_\tau$ and $\rho^*_\tau$
after splitting the signal into training and testing data. The HMM
obtained from a signal automaton at time $n$ can also be passed as
input to the Baum-Welch or similar algorithm in order to make it
converge to a (local) maximum. The latter idea raises the following
question.  Are there any generic or, at least, application dependent
forms of $\kappa^*_\tau$, $\sigma^*_\tau$ and $\rho^*_\tau$ such that
the global maximum can be reached upon feeding the Baum-Welch
algorithm with the model?
\bibliography{signal}
\end{document}